\documentclass [aps,pra,twocolumn,superscriptaddress]{revtex4}
\usepackage{amsmath}
\usepackage{amssymb}
\usepackage{graphicx}
\usepackage[T1]{fontenc}
\usepackage{color}

\usepackage{soul}

\begin{document}

\title{Circular dichroism in nanoparticle helices as a template for assessing quantum-informed models in plasmonics}

\author{Christos~Tserkezis}
\email{ct@mci.sdu.dk}
\affiliation{Center for Nano Optics, University of Southern Denmark, Campusvej 55, DK-5230 Odense M, Denmark}
\author{A.~T.~Mina~Ye\c{s}ilyurt}
\affiliation{Leibniz Institute of Photonic Technology, Albert-Einstein Str. 9, 07745 Jena, Germany}
\author{Jer-Shing~Huang}
\affiliation{Leibniz Institute of Photonic Technology, Albert-Einstein Str. 9, 07745 Jena, Germany}
\affiliation{Research Center for Applied Sciences, Academia Sinica, Taipei 11529, Taiwan}
\affiliation{Department of Electrophysics, National Chiao Tung University, Hsinchu 30010, Taiwan}
\author{N.~Asger~Mortensen}
\email{asger@mailaps.org}
\affiliation{Center for Nano Optics, University of Southern Denmark, Campusvej 55, DK-5230 Odense M, Denmark}
\affiliation{Danish Institute for Advanced Study, University of Southern Denmark, Campusvej 55, DK-5230 Odense M, Denmark}

\begin{abstract}
As characteristic lengths in plasmonics rapidly approach the sub-nm regime, quantum-informed models that can capture those aspects of the quantum nature of the electron gas that are not accessible by the standard approximations of classical electrodynamics, or even go beyond the free-electron description, become increasingly more important. Here we propose a template for comparing and validating the predictions of such models, through the circular dichroism signal of a metallic nanoparticle helix. For illustration purposes, we compare three widely used models, each dominant at different nanoparticle separations and governed by its own physical mechanism, namely the hydrodynamic Drude model, the generalised nonlocal optical response theory, and the quantum-corrected model for tunnelling. Our calculations show that indeed, each case is characterised by a fundamentally distinctive response, always dissimilar to the predictions of the local optical response approximation of classical electrodynamics, dominated by a model-sensitive absorptive double-peak feature. In circular dichroism spectra, the striking differences between models manifest themselves as easily traceable sign changes rather than neighbouring absorption peaks, thus overcoming experimental resolution limitations and enabling efficient evaluation of the relevance, validity, and range of applicability of quantum-informed theories for extreme-nanoscale plasmonics.
\end{abstract}
\maketitle


\section{Introduction}\label{Sec:Intro}

Plasmonics has experienced a drastic paradigm shift during the past decade, with rapid advances in nanofabrication and characterisation allowing metallic nanostructures to enter the quantum regime~\cite{tame_natphys9,zhu_natcom7,bozhevolnyi_nanophot6,Bozhevolnyi,tserkezis_ijmpb31_0}, defined by components of just a few nm in size~\cite{palpant_prb57,berciaud_nl5,scholl_nat483,raza_natcom6} and separations of only a few \AA~\cite{duan_nl12,kern_nl12}. In such situations, quantum effects like nonlocal screening~\cite{abajo_jpcc112,raza_jpcm27}, surface-enhanced Landau damping~\cite{shahbazyan_prb94,khurgin_acsphoton4}, electron spill-out~\cite{teperik_prl110,toscano_natcom6} and tunnelling~\cite{savage_nat491,scholl_nl13,tan_sci343,lerch_ijmpb31}, which cannot be captured by the common local response approximation (LRA) of classical electrodynamics, quickly become important and dominate the optical response. To deal with this issue, and overcome the limitations of the common free-electron-gas description in theoretical studies, a variety of quantum-informed models has been proposed.

Benefiting from the enormous experience acquired over the decades through studies of metals in solid-state physics, models based on screening and the  resulting nonlocal dielectric function~\cite{Ashcroft}, the $d$-parameter formalism of Feibelman~\cite{feibelman_progsurfsci12}, or (time-dependent) density functional theory (TD-DFT)~\cite{lang_prb1,Marques,varas_nanophot5,zhang_ijmpb31} have been developed. For instance, the hydrodynamic Drude model (HDM), already introduced in the 1930s~\cite{bloch_zphys81}, keeps reemerging in various forms due to its success in reproducing screening effects in noble metals~\cite{ruppin_prl31,fuchs_prb35,leung_prb42,mcmahon_jpcc114,raza_prb84,david_acsnano8,trugler_ijmpb31}, and approaches to extend its applicability are being proposed~\cite{toscano_natcom6,luo_prl111,ciraci_prb95}. One such extension, that also accounts in an efficient way for Landau damping, is provided by the generalised nonlocal optical response (GNOR) theory~\cite{mortensen_natcom5}. Other approaches to tackle Landau damping are typically based on a modified, quantum-mechanically obtained damping rate~\cite{shahbazyan_prb94,khurgin_acsphoton4}. On the other hand, the quantum-corrected model (QCM)~\cite{esteban_natcom3} has proven efficient in treating plasmonic dimers with gap distances in the quantum-tunnelling regime~\cite{savage_nat491,scholl_nl13} when the sizes involved render full TD-DFT calculations prohibitive. Finally, $d$-parameter approaches~\cite{yan_prl115,christensen_prl118} have attracted much interest recently, as they introduce the induced charges and currents of a quantum-mechanical calculation into an otherwise classical computation, accounting in principle for all aspects of quantum predictions relevant to plasmonics.

To move through this diversity of complementing -- or sometimes contradicting -- approaches, a robust test bed for assessing their validity and pertinence is required. Here we show that a promising such template can be found in chiral metallic nanoparticle (NP) helices~\cite{fan_jpcc115,kuzyk_nat483}, whose optical response is dominated by a double-peak absorption resonance in the visible. By calculating absorption spectra of such helices for left- and right-circularly polarised (LCP and RCP) incident light, we show that different models produce fundamentally different circular dichroism (CD) spectra, characterised by a sign change that can be traced even when the two individual absorption peaks are hard to resolve. In particular, we compare HDM, GNOR and QCM to the standard LRA, and explain how gradually decreasing the distance between NPs in a helical arrangement allows to monitor the transition from one model's predominance to the other's, through distinct, experimentally resolvable spectral features that are significantly different in each case, in accordance with the particular physics governing each model.

\section{Theoretical methods}\label{Sec:models}

A fundamental factor to the rapid growth of plasmonics has been that, in the vast majority of situations, the theoretical description is extremely efficient within classical electrodynamics, with the metal adequately described by a \emph{local}, usually scalar dielectric function $\varepsilon$. The main assumption of LRA is that, in the linear regime, the displacement vector $\mathbf{D}$ in a metallic NP is simply proportional to the externally applied electric field $\mathbf{E}$~\cite{Bohren},
\begin{equation}\label{Eq:localD}
\mathbf{D} (\mathbf{r}, \omega) = \varepsilon_{0} \varepsilon (\omega) \mathbf{E} (\mathbf{r}, \omega)~,
\end{equation}
where $\varepsilon_{0}$ is the vacuum permittivity, and $\varepsilon$ depends only on the angular frequency $\omega$ (dispersive medium), but not on the position $\mathbf{r}$. For simple free-electron metals $\varepsilon$ follows the Drude model~\cite{Ashcroft},
\begin{equation}\label{Eq:Drude}
\varepsilon = \varepsilon_{\infty} - \frac{\omega_{\mathrm{p}}^{2}}{({\omega^{2} + \mathrm {i} \omega \gamma)}}~,
\end{equation}
where $\omega_{\mathrm{p}}$ is the plasmon frequency, $\gamma$ is the damping rate, and $\varepsilon_{\infty}$ the contribution of bound electrons. In noble metals, the contribution from interband transitions is typically included in $\varepsilon_{\infty}$, either through fitting Eq.~(\ref{Eq:Drude}) to measured data~\cite{johnson_prb6}, or by subtracting the Drude part from the experimental dielectric function~\cite{abajo_jpcc112}. With these assumptions, one just needs to solve the wave equation for the electric field,
\begin{equation}\label{Eq:WaveLRA}
\nabla \times \nabla \times \mathbf{E} (\mathbf{r}, \omega) = \left(\frac{\omega}{c}\right)^{2} \varepsilon (\omega) \mathbf{E} (\mathbf{r}, \omega)~,
\end{equation}
subject to the appropriate boundary conditions. In LRA, a hard wall boundary condition is usually adopted, implying that the induced charges are exactly localised at the metal-dielectric interface, explicitly excluding the possibility of spill-out, while the electron gas at the NP interior is assumed incompressible.

The assumption of locality starts to become problematic when NP sizes, or their separations in plasmonic aggregates, are comparable to the electron mean free path. Going one step backwards, a more fundamental (still in the linear regime) constitutive relation is
\begin{equation}\label{Eq:nonlocalD}
\mathbf{D} (\mathbf{r}, \omega) = \varepsilon_{0} \int \mathrm{d} \mathbf{r}' \varepsilon (\mathbf{r}, \mathbf{r}', \omega) \mathbf{E} (\mathbf{r}', \omega)~,
\end{equation}
where the displacement at position $\mathbf{r}$ might depend on the electric field at positions $\mathbf{r}'$. In a homogeneous medium the spatial dependence becomes $\mathbf{r}-\mathbf{r}'$, simplifying the Fourier-transformed Eq.~(\ref{Eq:nonlocalD}) to $\mathbf{D} (\mathbf{k}, \omega) = \varepsilon_{0} \varepsilon(\mathbf{k}, \omega) \mathbf{E} (\mathbf{k}, \omega)$, where $\mathbf{k}$ is the wavevector. To proceed one needs an explicit form for $\varepsilon(\mathbf{k}, \omega)$, such as those obtained in the theory of screening~\cite{Ashcroft}. An efficient description is provided by HDM, where the equation of motion (at time $t$) for an electron moving with velocity $\mathbf{v} (\mathbf{r}, t)$ in a metal with electron density $n (\mathbf{r}, t)$ is given by
\begin{equation}\label{Eq:HydroMotion}
\left[\frac{\partial}{\partial t} + \mathbf{v} \cdot \nabla \right] \mathbf{v} = - \frac{e}{m} \left(\mathbf{E} + \mathbf{v} \times \mathbf{B} \right)  - \frac{1}{m} \nabla \frac{\delta G[n]}{\delta n} - \gamma \mathbf{v}~.
\end{equation}
In the above, $e$ is the (positive) electron charge and $m$ its mass; $\mathbf{B}$ is the magnetic field, so that the first term on the right-hand side corresponds to the Lorentz force, while the last term describes damping. Finally, the functional $G$ takes into account the internal kinetic energy of the electron gas, together with the exchange and correlation contributions. In its simplest form, the Thomas--Fermi approximation, only the kinetic energy is accounted for, and the second term becomes $(\beta^{2}/n) \nabla n$, where, in the high-frequency limit, $\beta^{2} = 3 v_{\mathrm{F}}^{2}/5$ ($v_{\mathrm{F}}$ being the Fermi velocity of the metal). This hydrodynamic pressure term accounts for the finite compressibility and the fact that electrons are fermions obeying the exclusion principle~\cite{Griffiths}. Eq.~(\ref{Eq:HydroMotion}), combined with the continuity equation, $\frac{\partial}{\partial t} n = - \nabla \cdot (n \mathbf{v})$, can be solved assuming a small deviation from the equilibrium electron density $n_{0}$ to produce the system of coupled electromagnetic (EM) equations~\cite{raza_jpcm27}
\begin{eqnarray}\label{Eq:CoupledHydro}
&& \nabla \times \nabla \times \mathbf{E} (\mathbf{r}, \omega) = \left(\frac{\omega}{c}\right)^{2} \varepsilon_{\infty} \mathbf{E} (\mathbf{r}, \omega) + \mathrm{i} \omega \mu_{0} \mathbf{J} (\mathbf{r}, \omega)~, \nonumber\\
&& \frac{\beta^{2}}{\omega \left(\omega + \mathrm{i} \gamma \right)} \nabla \left[\nabla \cdot \mathbf{J} (\mathbf{r}, \omega) \right] + \mathbf{J} (\mathbf{r}, \omega) = \sigma \mathbf{E} (\mathbf{r}, \omega)~,
\end{eqnarray}
where $\mu_{0}$ is the vacuum permeability, $\mathbf{J} (\mathbf{r}, \omega) = -e n_{0} \mathbf{v} (\mathbf{r}, \omega)$ is the induced current density, and $\sigma = \varepsilon_{0} \mathrm{i} \omega_{\mathrm{p}}^{2}/[\omega (\omega + \mathrm{i} \gamma)]$ is the Drude conductivity. While exact analytic solutions can be found for spherical and cylindrical NPs~\cite{ruppin_prl31,raza_prb84}, for arbitrary geometries one solves Eqs.~(\ref{Eq:CoupledHydro}) numerically. Here we use a commercial finite-element method (FEM) solver (Comsol Multiphysics 5.1)~\cite{toscano_oex20}, assuming the additional boundary condition $\mathbf{J} \cdot \hat{\mathbf{n}} = 0$ (where $\hat{\mathbf{n}}$ is the unit vector normal to the metal-dielectric interface), which implies that the electron density vanishes abruptly at the interface (no spill-out). This condition can be relaxed self-consistently, by adding exchange-correlation terms to the functional $G[n]$ of Eq.~(\ref{Eq:HydroMotion})~\cite{toscano_natcom6}.

With HDM describing electron convection, one might ask how the optical response is affected by diffusive currents. In the presence of diffusion the continuity equation becomes
\begin{equation}\label{Eq:ContinuityDiffusion}
-\mathrm{i} \omega e n(\mathbf{r}, \omega) = D \nabla^{2}\left[e n(\mathbf{r}, \omega)\right] + \nabla \cdot \left[-e n_{0}  \mathbf{v}(\mathbf{r}, \omega)\right]~,
\end{equation}
where $D$ is the diffusion constant. Combining this with the diffusive version of Fick's law,
\begin{equation}\label{Eq:Fick}
\mathbf{J}(\mathbf{r}, \omega) = -e n_{0} \mathbf{v}(\mathbf{r}, \omega) + e D \nabla n(\mathbf{r}, \omega)~,
\end{equation}
simply modifies the second of Eqs.~(\ref{Eq:CoupledHydro}) by adding the term $-\mathrm{i} D/\omega$ to the prefactor of $\nabla \left[\nabla \cdot \mathbf{J} (\mathbf{r}, \omega) \right]$; the solving methodologies developed for HDM apply thus immediately. Using the Boltzman equation of motion as the starting point, it can be shown that diffusion is the main \textit{low-frequency} contribution in the bulk, and becomes negligible at optical frequencies. Consequently, if a diffusion term is to be included, it must describe processes at the surface of the metal. Indeed, comparison with TD-DFT and modified-damping models shows that GNOR accounts efficiently for surface-enhanced Landau damping~\cite{tserkezis_ijmpb31}.

For NP separations of just few {\AA}, a prominent feature left out of the hydrodynamic treatment (by assuming that $\mathbf{J} \cdot \hat{\mathbf{n}} = 0$ at the metal surface) is the probability of direct electron tunnelling between NPs~\cite{savage_nat491,scholl_nl13}. This is where QCM~\cite{esteban_natcom3,esteban_faraday178} comes into play, to account for the infeasibility of fully quantum mechanical calculations for realistic NP sizes, where millions of electrons are involved. The model starts by calculating (typically with TD-DFT) the electron tunnelling probability in a nanogap (e.g. two flat interfaces), $T(\Omega, l)$, as a function of energy $\Omega$ and position in the gap $l$. This in turn is translated into a distance-dependent dc conductivity through \cite{esteban_natcom3}
\begin{equation}\label{Eq:QCMconductivity}
\sigma_{0}(l) = \frac{l}{2\pi^{2}} \int_{0}^{\Omega_{\mathrm{F}}} \mathrm{d}\Omega \; T(\Omega, l)~,
\end{equation}
where $\Omega_{\mathrm{F}}$ is the Fermi energy of the metal. A tunnelling damping rate $\gamma_{\mathrm{g}}$ is then obtained from 
\begin{equation}\label{Eq:QCMGamma}
\gamma_{\mathrm{g}} = \frac{\omega_{\mathrm{g}}^{2}}{4\pi \sigma_{0}(l)}~, 
\end{equation}
where the gap plasma frequency $\omega_{\mathrm{g}}$ is assumed equal to that of the bulk metal, $\omega_{\mathrm{p}}$. These two parameters, $\omega_{\mathrm{g}}$ and $\gamma_{\mathrm{g}}$, are finally introduced into a classical EM calculation, to model the Drude permittivity of a bridge connecting the metallic components [see schematics in Fig.~\ref{fig5}(a)]. This approach has been employed to mimic the role of tunnelling-mediated dissipation in different geometries, such as NP aggregates~\cite{hohenester_prb91} or nanomatryoshkas of alternating metal and dielectric layers~\cite{zapata_oex23}.

\section{Optical response of NP helices}\label{Sec:LRA}

\begin{figure}[h]
\centerline{\includegraphics*[width=1\columnwidth]{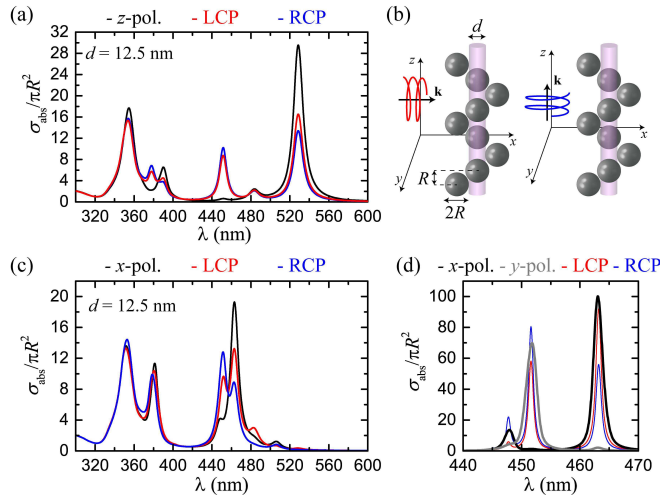}}
\caption{(a) Absorption cross section ($\sigma_{\mathrm{abs}}$, normalised to the geometrical cross section of a single NP), calculated within LRA for a linearly ($z$)- or circularly-polarised incident plane wave propagating along the $x$ axis, for a helix of 9 silver NPs ($R = 5$\;nm) revolving around a supporting pillar of diameter $d = 12.5$\;nm, as shown schematically in (b) (left-hand schematic). Black, red, and blue lines correspond to $z$-polarised, LCP, and RCP light, respectively. (c) Same as in (a), for propagation along the helix ($z$) axis [right-hand schematics in (b)]. (d) Zoom in the 440--480\;nm spectral window of (c), for silver NPs described by a dielectric function whose imaginary part is artificially reduced by 90\%. The spectra for linearly $x$- and $y$-polarised light are shown by a thick black and grey line respectively, while LCP (RCP) spectra are represented by a thin red (blue) line.}
\label{fig1}
\end{figure}

The quantum-informed models described in the previous section will be employed to explore the optical response of metallic NP helices. The helices considered in the largest part of the paper consist of 9 silver spheres with radius $R = 5$\;nm, described by the experimental dielectric function of Johnson and Christy~\cite{johnson_prb6}, in air. The spheres revolve around the $z$ axis by $\pi/2$ angle steps, so that 9 NPs produce two full revolutions, as shown schematically in Fig.~\ref{fig1}(b). While it was not modelled in our simulations, a supporting pillar of diameter $d$ and height 40\;nm is implied and shown in the schematics: such a pillar is usually produced with DNA-origami nanofabrication techniques~\cite{kuzyk_acsphoton5}. Our choice of NP size serves therefore a dual purpose: it is small enough to ensure that quantum effects will be relevant even for non-interacting NPs, while being the typical size supported by DNA-origami pillars~\cite{zhou_acr50}. The centre of each NP is vertically shifted by $R$, while $d$ can vary from 14 to 12.5\;nm, producing surface-to-surface NP distances of 3.5 to 0.3\;nm. The system is illuminated by circularly polarised light propagating either along or normally to the helix axis ($z$ axis).

In Fig.~\ref{fig1}(a) we plot the calculated absorption cross section, $\sigma_{\mathrm{abs}}$, normalised to the geometrical cross section of a single NP,  as a function of wavelength $\lambda$, for a helix revolving around a pillar with $d = 12.5$\;nm, which reduces the interparticle gap to just 0.3\;nm. For the strong interaction resulting from such narrow gaps, the long-wavelength resonances in the spectra can be understood in view of the embedded-chain model as collective chain modes in the direction of the electric field of the incident plane wave~\cite{taylor_jpcc120}, while short-wavelength resonances are associated with single-NP and higher-order hybrid modes. In this respect, when the incident plane wave propagates normally to the helix axis (here along the $x$ axis), the main difference between a linearly, $z$-polarised wave with electric field $\mathbf{E} = E_{0}\;\widehat{\mathbf{z}}$ ($E_{0}$ being the amplitude of the electric field) and an L(R)CP wave [taken as $\mathbf{E} = (E_{0}/\sqrt{2}) (\widehat{\mathbf{z}} \pm \mathrm{i} \widehat{\mathbf{y}})$ here] regards the excitation of short, kinked chains along the $y$ axis in the latter case, which manifests itself as an intense absorption peak around 450\;nm in the spectra. Naturally, since the system lacks mirror symmetry, LCP and RCP illumination produces different intensities of the absorption peaks, and thus a strong CD signal, in agreement with the results of Ref.~\cite{fan_jpcc115} (emphasis will be placed on CD spectra in the next section, where the different quantum-informed models are compared).

\begin{figure}[h]
\centerline{\includegraphics*[width=0.9\columnwidth]{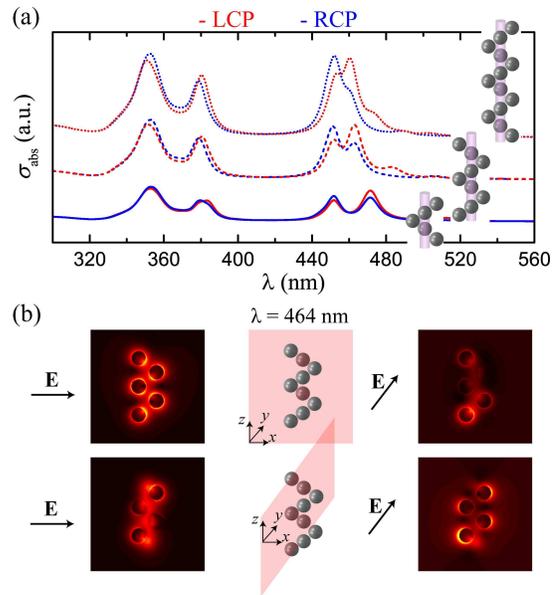}}
\caption{(a) Normalised absorption spectra (vertically shifted for clarity) for the Ag NP helices schematically shown on the right. From bottom to top, LCP (red) and RCP (blue) absorption spectra are calculated for helices of 5 (one full revolution, solid lines), 9 (two revolutions, dashed lines), and 13 NPs (three revolutions, dotted lines). (b) Near-field enhancement ($|\mathbf{E}|/E_{0}$) plots for the 9-NP helices of Fig.~\ref{fig1}(c), at $\lambda = 464$\;nm. Left- and right-hand panels correspond to $x$- and $y$-linearly polarised incident light, respectively. In the top row the field is plotted in the $x-z~(y=0)$ plane, and in the bottom row in the $y-z~(x=0)$ plane, as shown in the schematics in the middle panel.}\label{fig2}
\end{figure}

A far more interesting optical response is obtained for propagation along the helix axis, as can be seen in Fig.~\ref{fig1}(c). In this case, instead of a single long-wavelength peak as in Fig.~\ref{fig1}(a), a double peak appears in the spectral window of 440--480\;nm, centred at 452 and 464\;nm, with different intensity for each branch under LCP and RCP illumination [$\mathbf{E} = (E_{0}/\sqrt{2}) (\widehat{\mathbf{x}} \pm \mathrm{i} \widehat{\mathbf{y}})$ for L(R)CP light]. An additional, higher-order hybrid chain mode is excited at 448\;nm under $x$ polarisation, but as it always appears just as a shoulder at the high-energy end of the fundamental chain resonance~\cite{tserkezis_ppsc31}, its presence does not significantly affect the doublet of interest here. To understand the origin of this doublet, we plot in Fig.~\ref{fig1}(d) the spectra for linear polarisation along the $x$ or $y$ axis (thick black and grey line, respectively), with absorptive losses artificially reduced by 90\%, by manually modifying the imaginary part of the dielectric function of silver. It is clear that the 452\;nm mode is only excited by $y$ polarisation, while the 464\;nm only by $x$ polarisation. These resonances can therefore be understood as chain modes of embedded chains growing along the $x$ or $y$ axis, depending on the polarisation, appearing at different wavelengths due to the finite size of the helix. In an infinite helix, the number of embedded chains along both axes is the same, and the two peaks coincide. This can be verified by calculations for shorter (5 NPs, one revolution) and longer (13 NPs, three revolutions) helices, with the spectral split gradually closing, from 21 to 8\;nm, as shown in Fig.~\ref{fig2}(a). This interpretation of the modes sustained by such helices is further supported by the near-field profiles of Fig.~\ref{fig2}(b), where the highest field intensities are obtained at the gaps along the corresponding effective chain for each incident linear ($x$ or $y$) polarisation.

\begin{table}[ht]
\centering \caption{Character table of the $C_{2}$ point group.}\label{Table1}
\begin{tabular}{|c|c c|}
\hline 
~ & $\quad \mathcal{E} \quad$ & $\quad \mathcal{C}_{2x} \quad$ \\
\hline 
$Q_{1}$ & 1 & 1 \\
$Q_{2}$ & 1 & -1 \\
\hline
\end{tabular}
\end{table}

To further analyse the spectral doublet, we turn to group theory. Since the only symmetry operations $\widehat{P}$ leaving the finite helices unaltered are identity, $\mathcal{E}$, and a $\pi$ rotation about the $x$ axis, $\mathcal{C}_{2x}$, the appropriate point group is $C_{2}$, whose character ($\chi$) table is given in Table~\ref{Table1}~\cite{Cornwell}. All modes must have the symmetry of the irreducible representations of this group, namely $Q_{1}$ and $Q_{2}$, while a mode is excited only if the electric field of the incident light has a nonvanishing projection onto the appropriate subspace. For the $C_{2}$ group the projection operator is
\begin{equation}\label{Eq:projection}
\mathcal{P}^{(\Gamma)} = \frac{1}{2} \sum_{\Gamma} \chi^{\ast} (\Gamma) \widehat{P} \mathbf{E}~,
\end{equation}
where $\Gamma = \{\mathcal{E}, \mathcal{C}_{2x}\}$ is the set of group operations. Since a $\pi$ rotation about the $x$ axis leads to the following transformations: $x \rightarrow x$, $y \rightarrow -y$, $z \rightarrow -z$, $\mathbf{k} \rightarrow -\mathbf{k}$, it is straightforward to show that an $x$-polarised plane wave projects only onto $Q_{1}$, while a $y$-polarised plane wave only onto $Q_{2}$, in agreement with the absorption spectra of Fig.~\ref{fig1}(d). For circularly polarised light, the electric field has both $x$ and $y$ components and both modes are efficiently excited.

The reported optical response is rather robust against manufacturing deviations, facilitating realisation of the proposed experiment. We have already discussed how the number of NPs forming the helix affects the measured spectra, and concluded that short chains are much more preferable than longer ones. The number of helical revolutions is in fact the factor that calls for most attention. In the Supporting Information we present spectra for intermediate NP numbers (incomplete revolutions), increasing from 5 to 9 (Fig.~S1). It is shown that in those cases the spectral doublet is always efficiently excited. We also explore the role of the revolution angle, by considering $\pi/3$ steps (Fig.~S2), which are often more feasible with DNA-origami methods~\cite{kuzyk_nat483}. Similarly, the spectra are little affected by variations in the NP size and interparticle gap (we modelled up to $\pm 10 \%$ size variations in our calculations) within the same helix (Fig.~S3). On the other hand, intrinsic losses in the chosen material can be important, as we show through the corresponding spectra for helices of 9 gold NPs (Fig.~S4). There, the resonances are not only shifted to longer wavelengths, but also broadened so much that the doublet is not distinguishable in the absorption spectra; it might survive as a sign change in CD for some chain lengths. Finally, it is important to notice that in the following analysis, and in any experimental realisation, the exact values of resonance shifts or CD measurements are not important, and only the qualitative features of the spectra matter.

\section{Quantum corrections in the helix response}\label{Sec:Quantum}

\begin{figure}[h]
\centerline{\includegraphics*[width=0.9\columnwidth]{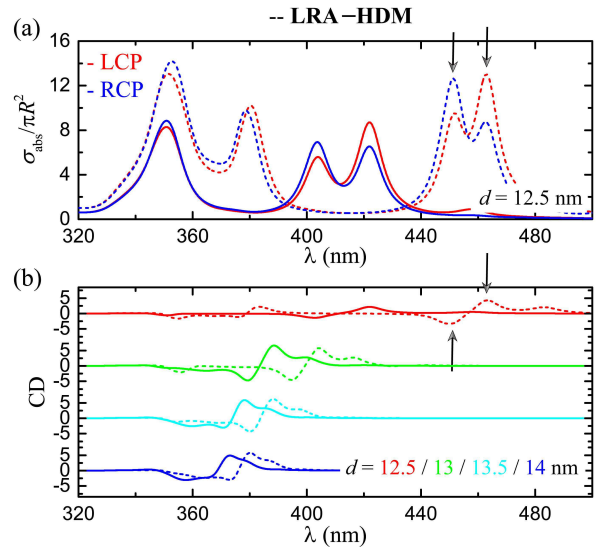}}
\caption{(a) Normalised absorption cross section for the helix of Fig.~\ref{fig1}(b) (right-hand schematic), for LCP (red) and RCP (blue) incident light, within the LRA and HDM description (dashed and solid lines respectively). (b) CD spectra for the two models, when increasing the interparticle gap through $d = 12.5, 13, 13.5, 14$\;nm (red, green, light-blue and blue lines respectively, corresponding to gaps of 0.31\;nm, 0.93\;nm, 1.55\;nm, and 2.18\;nm). The arrows in both panels trace the transition from the double-peak feature in absorption to the positive-negative signal in CD, highlighted here for the $d = 12.5$\;nm case (LRA calculation).}\label{fig3}
\end{figure}

In the remaining of the paper we explore how the absorption doublet discussed in the previous section, and the resulting CD signal (calculated here as the difference of absorption cross sections for LCP and RCP incident light), behave within the three different quantum-informed models presented above. We start with the HDM calculation, whose screening mechanism is expected to dominate for larger NP separations. We use $v_{\mathrm{F}} = 1.39 \times 10^{6}$\;m\;s$^{-1}$ for the Fermi velocity of silver, while $\varepsilon_{\infty}$ is obtained by subtracting from the experimental dielectric function a Drude permittivity with $\hbar \omega_{\mathrm{p}} = 8.99$\;eV and $\hbar \gamma = 0.25$\;eV~\cite{tserkezis_scirep6}. In Fig.~\ref{fig3}(a) we compare the LRA absorption spectra of Fig.~\ref{fig1}(c) (dashed lines) with those obtained by HDM (solid lines). As expected, the main difference is a large blueshift of the modes due to screening, which is the characteristic of HDM: since the induced charges are ``pushed'' inwards, the NPs behave as if they were effectively smaller, and therefore their separations larger, decreasing the strength of their interaction. This can be observed more clearly in CD spectra, through the transition from negative (short-wavelength branch) to positive (long-wavelength branch) peaks, as shown in Fig.~\ref{fig3}(b). The fingerprint of the blueshifting doublet is clearly visible even for $d = 14$\;nm, corresponding to an interparticle gap as wide as 2.2\;nm, for which individual absorption peaks are hard to resolve, as they nearly coincide in wavelength with the strong single-NP resonance (chain modes are only very weakly excited).

\begin{figure}[h]
\centerline{\includegraphics*[width=0.9\columnwidth]{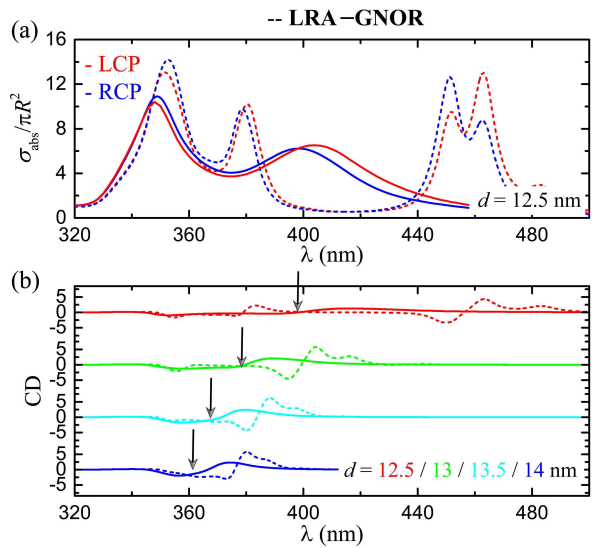}}
\caption{(a) Normalised absorption cross section for the helix of Fig.~\ref{fig1}(b) (right-hand schematic), for LCP (red) and RCP (blue) incident light, within the LRA and GNOR description (dashed and solid lines respectively). (b) CD spectra for the two models, when increasing the interparticle gap through $d = 12.5, 13, 13.5, 14$\;nm (red, green, light-blue and blue lines respectively, corresponding to gaps of 0.31\;nm, 0.93\;nm, 1.55\;nm, and 2.18\;nm). The arrows indicate the wavelength at which the CD signal changes sign, in the region of the double resonance of interest.}\label{fig4}
\end{figure}

It is now interesting to apply the same analysis for the GNOR model. To that end, we perform FEM simulations to solve the coupled Eqs.~(\ref{Eq:CoupledHydro}) (as modified for GNOR), with a diffusion constant $D = 2.684 \times 10^{-4}$\;m$^{2}$\;s$^{-1}$, a value which allows direct comparison with previous studies of silver NP dimers and chains~\cite{tserkezis_prb96}. Since GNOR is specifically designed to account for increased absorptive losses when NP sizes and distances are small, and CD is by definition an observable depending on absorption, one expects substantial differences from the response predicted by LRA. Indeed, as seen in Fig.~\ref{fig4}(a), GNOR introduces such a degree of plasmon damping and broadening, that the two separate peaks of interest merge into a single broad resonance (blueshifted with respect to LRA due to nonlocal screening, already accounted for with HDM). In this case, the only way to trace the double resonance of the helix chain modes is through its CD signal, which, even in this case, retains the characteristic sign change. This is shown in Fig.~\ref{fig4}(b), for the same four NP separations as in Fig.~\ref{fig3}(b). One should notice, however, that broadening of the modes could have many different origins, and is not necessarily an indication of Landau damping (see also the spectra for gold NP helices in the Supporting Information, Fig.~S4). Consequently, in an experiment, one should take care to choose materials with low intrinsic loss, and NPs as smooth and with as homogeneous size distributions as possible.	

\begin{figure}[h]
\centerline{\includegraphics*[width=0.9\columnwidth]{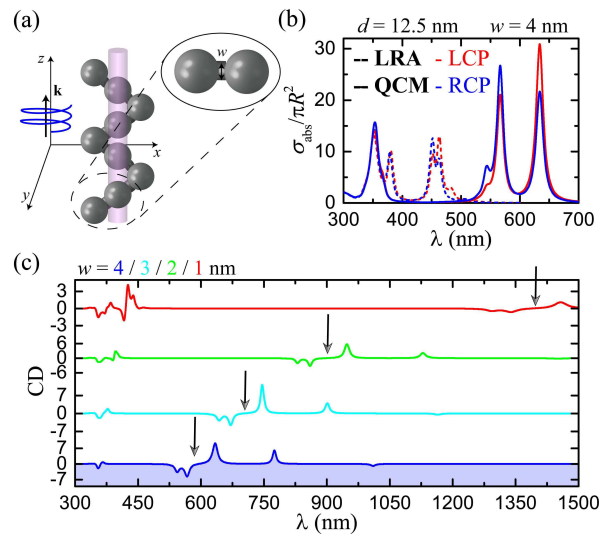}}
\caption{(a) Schematic of the bridged NP helices used for a qualitative implementation of QCM: cylindrical bridges of diameter $w$, described by a Drude dielectric function with values for $\omega_{\mathrm{g}}$ and $\gamma_{\mathrm{g}}$ appropriate for silver, connect the NPs. (b) Absorption spectra for LCP and RCP (red and blue lines) incident light, calculated for the helix of Fig.~\ref{fig1}(b) within LRA (dashed lines) and QCM (solid lines) ($w = 4$\;nm). (c) CD spectra in QCM for varying $w$, from $w = 1$ to 4\;nm (from top to bottom). The shaded line for $w = 4$\;nm represents the spectra for a higher value of $\hbar \gamma_{\mathrm{g}} = 0.04$\;eV, corresponding to lower tunnelling conductivities in the gap. The arrows indicate the wavelength at which the CD signal changes sign in the region of the double resonance of interest.}\label{fig5}
\end{figure}

Finally, for gaps as narrow as a few \AA, it is important to consider the possibility of direct electron tunnelling from one NP to another~\cite{savage_nat491,yan_prl115}. As described in Sec.~\ref{Sec:models}, QCM seeks to mimic this regime through a (usually multilayered) conductive junction, introduced in a classical EM calculation to connect neighbouring NPs, as shown schematically in Fig.~\ref{fig5}(a). For a strict, quantitative description, the conductivity of such a bridge must be obtained from quantum mechanical calculations (for reduced NP sizes that make it feasible), typically with TD-DFT~\cite{esteban_natcom3}. Nevertheless, since we are only interested in a qualitative description here, we use a homogeneous cylinder described by a Drude model, with $\hbar \omega_{\mathrm{g}} = 8.99$\;eV and $\hbar \gamma_{\mathrm{g}} = 0.025$\;eV (but with $\varepsilon_{\infty} = 1$, as in the previous calculations. Omitting the multilayered structure of the bridge can be compensated by considering different bridge diameters, which, through facilitating or hindering charge transfer between NPs, tune the optical response of the entire structure. On the other hand, the tunnelling conductivity might in fact be smaller than that of the bulk metal, and therefore, the Drude damping rate might be larger, according to Eq.~(\ref{Eq:QCMGamma}), leading to broadening of the modes. It turns out that in our case this broadening is rather small, as shown by the shaded blue line ($w = 4$\;nm) in Fig.~\ref{fig5}(c) which corresponds to $\hbar \gamma_{\mathrm{g}} = 0.04$\;eV, and is practically indistinguishable from the $\hbar \gamma_{\mathrm{g}} = 0.025$\;eV case (solid line). In the Supporting Information we show how the optical response of silver NP dimers is affected by these two parameters (Fig.~S5).

Fig.~\ref{fig5}(b) shows the absorption spectra obtained with QCM when Drude bridges of diameter $w = 4$\;nm connect each NP in the helix with its nearest neighbours. Once such junctions are created, threaded chain plasmons (TCPs) are excited in the infrared~\cite{herrmann_natcom5,tserkezis_oex22}, at longer wavelengths for thinner bridges in which charge transfer is more hindered, as can also be seen in the CD spectra of Fig.~\ref{fig5}(c). It is important to notice that these modes are not of the same nature as the chain modes discussed above and just redshifted, but hybrid, chain/rod (or, in our case, chain/spiral) modes. The unthreaded chain modes become screened chain plasmons, getting rapidly damped and blushifting due to the screening from the field accumulated around the bridge~\cite{perez_njp13}. The two limiting cases in threaded chains are therefore i) very weak, deep-infrared modes for $w \rightarrow 0$ and ii) single spiral (threaded helix) modes for $w \rightarrow 2R$~\cite{tserkezis_oex22}. Nevertheless, if tunnelling prevails, the spectra are dominated by a strongly redshifted double peak, predicting yet another qualitatively different optical response, characteristic of QCM, the exact quantitative aspects of which (in terms of resonance shift and strength) depend on the precise TD-DFT input~\cite{esteban_natcom3}.

\section{Discussion}\label{Sec:discussion}

Having explored three different quantum-informed models, one might ask what the real optical response of the helix will be, and what kind of resonances one should expect in experiments. The truth however is that, before measuring, there is no real way to know which predictions are more accurate. It can be expected that for larger NP separations, screening will prevail and spectra closer to those discussed in relation to Fig.~\ref{fig3} will be obtained. On the other hand, for small separations, both damping and tunnelling could become important, leading to broad, merged absorption peaks (but probably still traceable in CD), both at shorter (screened chain plasmons) and at longer (TCPs) wavelengths. Such a combination of effects could be predicted by more elaborate models, e.g. $d$-parameter based ones. But the situation is actually reminiscent of the debate about nonlocality a few years ago. The classical HDM predicts invariably resonance blueshifts. It took \emph{ab initio} calculations~\cite{teperik_prl110}, which for small monomers and dimers were feasible, together with sensitive experiments~\cite{rainers_prl74}, to establish the fact that for good free-electron metals (e.g. Na; in noble metals spill-out is much less important due to d-electron screening) spill-out prevails over screening, thus leading to resonance redshifts. Similarly, only fine experiments (the number of electrons involved here make \emph{ab initio} calculations impossible) can tell what the actual optical response of the helices is, and in doing so they will most likely call for the extension of the existing, or the development of new quantum-informed models.

Finally, let us briefly comment on the feasibility of the proposed architectures. One way of fabricating advanced plasmonic devices that exhibit novel optical properties is utilising DNA origami \cite{liu_chemrev118}. Due to its programmability and specificity, DNA origami enables NP organisation at the sub-nm scale \cite{bidault_jacs130}, that is still a challenge for top-down techniques. For instance, a single stranded DNA could be folded by smaller single DNA strands, called ssDNAs, to build a pillar with a diameter 14-12.5\;nm. It has been shown that 10-15 nucleotides, approximately 1.3\;nm in length, are enough to provide this connection \cite{heck_acsphoton4}. The pillar would be robust enough to support metallic NPs of diameter 10\;nm \cite{kuzyk_nat483}, and NPs would stick onto their predefined locations by the use of capture strands protruding from the pillar. Hence, it should be possible to obtain interparticle gaps of 0.3-3.5\;nm. Another possibility could be directly growing NPs on a DNA template \cite{shemer_jacs128}. In the case of silver in particular, oxidation can also be an issue, and experiments should be performed as soon as carefully synthesised NPs are prepared \cite{scholl_nature483}, to reduce its influence \cite{weller_acsphoton3}.

\section{Conclusions}\label{Sec:conclusion}

In summary, we have shown that CD measurements in chiral chains of metallic NPs, experimentally feasible nowadays with DNA-origami techniques, serve as an excellent test bed for assessing quantum-informed models in plasmonics. By comparing HDM, GNOR, and QCM, we obtained three fundamentally different optical responses, each one characteristic of the specific model. The spectral doublet calculated within LRA shifts to the blue in HDM due to screening, merges into a single resonance due to broadening in GNOR, and strongly redshifts once tunnelling becomes important in QCM. CD measurements allow to monitor these changes simply through a change in sign. Even though our calculations only intend to qualitatively describe each model, we anticipate that quantitative differences should be large enough to be experimentally resolved, thus allowing to understand the limitations and range of validity of, in principle, any existing or new quantum-informed model.

\begin{acknowledgements}
We thank C. Wolff for discussions. N.~A.~M. is a VILLUM Investigator supported by VILLUM FONDEN (grant No. 16498). The Center for Nano Optics is financially supported by the University of Southern Denmark (SDU 2020 funding). Simulations were supported by the DeIC National HPC Centre, SDU.
\end{acknowledgements}

\end{document}